\documentclass[twocolumn,prl,showpacs]{revtex4}
\usepackage{graphicx}
\usepackage{dcolumn}
\usepackage{amsmath}
\usepackage{latexsym}

\begin{document}
\title{Stress relaxation in a perfect nanocrystal
by coherent ejection of lattice layers}  
\author{Abhishek Chaudhuri and Surajit Sengupta
}
\affiliation{
Satyendra Nath Bose National Centre for Basic Sciences, Block-JD, Sector-III, Salt Lake, Calcutta - 700098, India
}
\author{Madan Rao}
\affiliation{ Raman Research Institute, C.V. Raman Avenue, Bangalore 560080, 
India\\
National Centre for Biological Sciences-TIFR, GKVK Campus, Bellary Road, Bangalore 560036,
India}
\date{\today}
\begin{abstract}
We show that a small crystal trapped within a potential well and in contact 
with its own fluid, 
responds to large compressive stresses by a novel mechanism --- the transfer 
of complete lattice layers across the solid-fluid 
interface. Further, when the solid is impacted by a momentum impulse set 
up in the fluid, a coherently ejected lattice layer carries away 
a definite quantity of energy and momentum, resulting in a sharp peak in 
the calculated phonon absorption spectrum. Apart from its relevance to studies 
of stability 
and failure of small sized solids, such coherent nanospallation may be used to
make atomic wires or monolayer films.
\end{abstract} 
%\pacs{62.,64.,68.08.-p,68.65.-k}
\maketitle

\noindent
Solids subject to large uniaxial deformations, relieve stress either by the 
generation and mobility of dislocations\cite{haasen} and/or by the 
nucleation and growth of cracks\cite{haasen,marder}. What is the nature of 
stress relaxation when conditions are arranged such that these conventional 
mechanisms are suppressed? Nano-indentation experiments\cite{uzi} show that 
if small system size prevents the generation of dislocations\cite{debc}, 
solids respond 
to tensile forces by shedding atoms from the surface layer. In this Letter, we 
study equilibrium and dynamical aspects of this process in detail using 
Monte Carlo and molecular dynamics simulations\cite{daan} for a model 
system, analyzing our results in the light of 
existing theory\cite{landau-el,debc,landau}. Briefly, we discover that a 
small solid,
constrained to remain defect free by being, at all times, in contact 
with its own liquid (a situation easily realised using optical 
traps\cite{laser-trp}), responds to stress by exchanging surface atomic 
layers with the adjacent liquid. Impacting the solid with a momentum 
pulse\cite{spall} of sufficient strength dislodges an entire crystalline 
layer coherently, which travels
into the liquid as a distinct though short-lived
entity, with a lifetime determined by the fluid 
viscosity\cite{landau}. A curious feature of this process is  that weaker 
pulses {\em do not} dislodge partial layers, leading to a novel resonance 
phenomenon distinguished by a pronounced peak in sound 
absorption\cite{landau}.   
We believe that our work may be useful for undertanding the 
failure behaviour and sound and heat\cite{kapitza} absorption properties of 
nanostructures\cite{nanostuff,nems}. Since coherent scattering of momentum pulses 
occur over a narrow window of incident energies, this phenomenon may also be 
used as a detector/analyzer for weak acoustic shocks. Free standing, cleaved 
single atomic layers\cite{pnas} have recently been shown to posses interesting 
mechanical and electrical properties. Coherent spallation\cite{spall}
of nanocrystals, as discussed here, may be a practical way to produce 
such atomic layers or for making nanowires or  
nanosurface coatings\cite{sakhi} in the future.
\begin{figure}[t]
\begin{center}
\includegraphics[width=8.0cm]{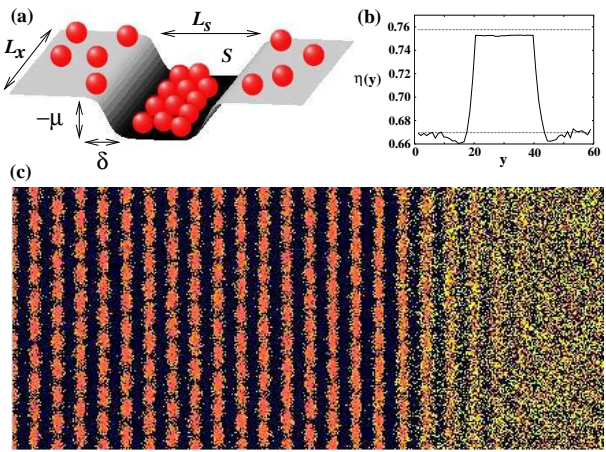}
\end{center}
\caption{(a) A 2d crystal of $N_s$ atoms confined to a central region ${\cal S}$ of area $A_s = L_x \times L_s$ by means of an optical trap
with a potential $\phi({\bf r}) = -\mu$ for ${\bf r} \in {\cal S}$ which increases sharply to zero elsewhere over a width $ \delta$. The trapped solid of density $\eta_{s}$ is in contact with its fluid of density $\eta_l < \eta_s$, such that the average density in the entire simulation cell of $N$ atoms occupying an area $A = L_x \times L_y$ is $\eta = \pi N/4 A$.  
The atoms interact with a   
{\em hard disk} potential\cite{debc,jaster,snb}  $V_{ij} = 0$ for 
$|{\bf r}_{ij}| > {\rm \sigma}$ and $V_{ij} = \infty$ for 
$|{\bf r}_{ij}| \leq {\rm \sigma}$, where ${\rm \sigma}$ is
the hard disk diameter which sets the scale of length and 
${\bf r}_{ij} = {\bf r}_j - {\bf r}_i$ the relative position vector of 
the particles. We have chosen $ \delta = \sigma/4$.   
We set the energy scale by $k_B T$.   
(b) Equilibrium behaviour for different 
$\mu$ at fixed $\eta = .699$ (density at freezing
$\eta_f = .706$\cite{jaster}) obtained by Monte Carlo simulations in the constant NAT  ensemble with periodic boundary conditions in both directions
($N=1200$ particles occupy an area $A = 22.78 \times 59.18$ with the 
solid occupying the central third of the cell of size $L_s = 19.73$).
The trap depth $\mu = 6$, supports an equilibrium solid of density $\eta_s = .753$ in contact with a fluid of density $\eta_{\ell} = .672$.
The closest-packed lines of the solid in ${\cal S}$ are parallel to the solid~-fluid interfaces 
which lie, at all times,
along the lines where $\phi(y) \to 0$. The density profile $\eta(y)$ coarse grained over strips of width $\sigma$ (averages taken over $10^3$ MC configurations each separated by $10^3$ MCS),
varies from $ \eta_{\ell}$ to $\eta_s$ as we move into ${\cal S}$. The horizontal lines are predictions of 
a simple free-volume based theory (see Fig. 2) for $\eta_s$ and $\eta_{\ell}$.   
(c) Superposition of particle configurations from the MC run in (b) 
showing a solid like order (red : high $\eta$) gradually vanishing into the fluid (yellow : low $\eta$) across a well defined solid-fluid interface. } 
\label{intfce}
\end{figure}

\noindent
We create a dislocation-free nanosolid by trapping a collection of
atoms in their crystalline state within a potential well of depth $-\mu$ over 
a finite region ${\cal S}$ placed in contact with its fluid (Fig.~1). Trapping 
of atoms such as alkali metals and noble gases may be achieved by optical and magneto-optical techniques \cite{laser-trp}, using laser powers ranging from $1-10^2$ mW. On the other hand, colloidal
solids\cite{colbook} may be 
manipulated using a number of optical techniques\cite{col-trp,laser-col}
or surface templates\cite{epitaxy}. 
\begin{figure}[t]
\begin{center}
\includegraphics[width=7.0cm]{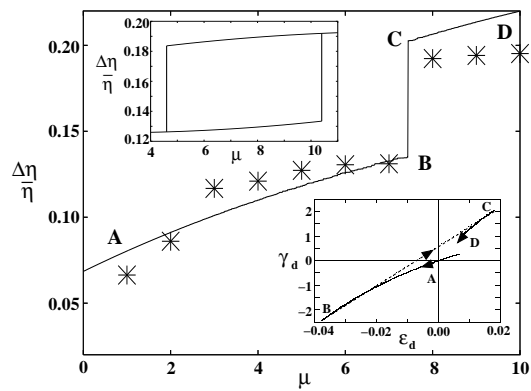}
\end{center}
\caption{ 
Plot of the equilibrium fractional density change $\Delta\eta/\eta$ as a function of 
$\mu$ (points (MC data), thick solid line (approximate theory)), showing discontinuous jump at $\mu \approx 8$. The MC data are obtained by averaging 
over $10^3$ configurations each separated by $10^3$ MCS, while the system is equilibrated for $>10^7$ MCS starting from a uniform fluid with 
density $\eta = .699$. The approximate theory is based on the assumption  that a change in $\mu$ produces a
{\em uniform} geometric strain $\varepsilon_d$ from a 
reference triangular lattice with the same number of atomic layers. 
%DO NOT delete the following sentence I think something like this is 
%needed 
The geometric strain $\varepsilon_d$ is an oscillatory function\cite{debc} of 
$L_s$, with an amplitude which decays as $1/L_s$.
%; vanishing 
%(by construction) for all $L_s$ which accomodate an integral number of 
%triangular lattice layers. 
The Helmholtz free energy of the harmonic solid 
is then given by
$f_s = f_{\Delta} + \frac{1}{2} K_{\Delta}\varepsilon_d^2$,
where $f_{\Delta}$ and $ K_{\Delta}$ are the free energy and Young's modulus
respectively, of an undistorted triangular lattice which may be obtained from 
simple free-volume theory\cite{debc,frevol}. 
Minimising the total free energy density of the fluid + solid regions,
$f = x \left[ f_s(\eta_s,L_s) - 4 \eta_s \mu/\pi\right] + (1-x) f_l(  \eta_{\ell})$
with the constraint 
$\eta = x \eta_s + (1-x)   \eta_{\ell}$, where $x$ is the area fraction occupied by ${\cal S}$ and $f_l(  \eta_{\ell})$ is 
the free energy of the hard disk fluid\cite{santos} produces the jump 
in $\Delta\eta(\mu)/\eta$. 
Inset (top) shows a cycle-averaged hysteresis loop
as $\mu$ is cycled at the rate of $.2$ per $10^6$ MCS.  (bottom) A plot of the 
tensile stress $\gamma_d$ against strain $\varepsilon_d$. The arrows show the 
behaviour of these quantities as $\mu$ is increased from the points marked 
A -- D. The corresponding points in the $\Delta\eta/\eta$ vs. $\mu$
plot is also marked for comparison. The state of stress in the solid jumps 
discontinuously from tensile to compressive from B\,$\rightarrow$\,C due to 
an increase in the number of solid layers by one accomplished by 
incorporating particles from the fluid. This transition is reversible
and the system relaxes from a state of compression to tension by ejecting 
this layer as $\mu$ is decreased.      
}
\label{hyst}
\vskip -.5cm
\end{figure}
\vskip 0.1cm

\noindent
To see that this trapped nanosolid is dislocation-free, we study the equilibrium behaviour of a model 2-dimensional (2d) nanosolid using a Monte 
Carlo simulation in the constant number ($N$), area ($A$) and temperature ($T$) 
ensemble with usual Metropolis moves\cite{debc,daan} using the Hamiltonian
${\cal H} = \sum_{ij} V_{ij} + \sum_{i}\phi({\bf r}_i)$, where $V_{ij}$ is the 
2-body potential and $\phi$ is the trap potential, here approximated as a rectangular potential well of 
depth $-\mu$ (Fig. 1). For 
numerical convenience we choose a system of hard disks (Fig. 1);
our main results trivially extend to particles interacting with any form of 
repulsive potential, or even when the interactions are augmented by a 
short-range attraction, provided we choose $\mu$ deeper than the depth of 
the attractive potential. While all our simulations are carried out in 2d, our qualitative results should extend to three dimensions (we shall use the generic words `layer' and `surface' to describe the 1-dimensional line of atoms).
The equilibrium density profile (Fig.1b) is obtained\cite{note} 
for different $\mu$ at fixed average density $\eta$. \\

\noindent
The `phase diagram', a plot of the density difference
across the interface $\Delta \eta/\eta \equiv (\eta_s -   \eta_{\ell})/\eta$ 
versus $\mu$ (Fig. 2), shows a sharp jump at $\mu \approx 8$. This transition is associated with an entire close-packed layer entering ${\cal S}$ thereby 
increasing the number of solid layers by one.
The resulting solid is a triangular lattice with a small 
rectangular distortion $\varepsilon_d(\eta_s,L_s)\,$\cite{debc}. The 
qualitative features of this phase diagram may be obtained by a simple 
thermodynamic theory (Fig. 2) with harmonic distortions of the solid, ignoring 
contributions from spatial variations of the density. We find that the jump in the fractional density difference is sensitive to $L_s$ and vanishes for large $L_s$ or $\delta$, the sharpness of the trap. \\

\noindent
Adiabatically cycling the trap depth $\mu$ across the jump obtains a 
sharp rectangular hysteresis loop; this indicates that `surface' steps (dislocation pairs) nucleated in the course of adding (or subtracting) a solid layer, have a vanishingly short lifetime.
Consistent with this we find that the jump in
$\Delta\eta(\mu)/\eta$ vanishes when the system is minimised at each $\mu$ 
with a constraint that the solid contains a single dislocation pair. %(Fig. 2). 
Interestingly, a dislocation pair forced initially into the bulk, rises 
to the solid-fluid interface due to a gain in strain energy\cite{landau-el}, 
where they form 
surface indentations flanked by kink-antikink pairs. The confining potential,
which prefers a flat interface, may remove these indentations
either by bending lattice layers or annealing the kink-antikink pair  
incorporating particles from the adjacent fluid. 
The second process always costs less energy and happens quickly.
The question is how quickly ?
\\

\noindent
To study the lifetime of the kink-antikink pairs (surface step), we resort to a
molecular dynamics (MD) simulation using a 
velocity Verlet algorithm\cite{daan}, with the unit of time
given by $\tau = \sqrt{m \sigma^2/k_B T}$, where  
$m (= 1)$ is the mass of the hard disks\cite{note-energy}. 
Using values of $m$ and $\sigma$ typical for atomic systems like Ar or 
Rb, $\tau \approx 1 {\rm ps}$. 
Starting with an equilibrium configuration at $\mu = 9.6$ (and $k_B T =1$) 
corresponding to a $22$-layer solid, we create a unit surface step of 
length $l$ by removing a few interfacial atoms and `quench' 
across the transition to $\mu=4.8$, where the $21$-layer solid is stable.
A free energy audit involving a bulk free energy gain $\Delta F L_s l$, 
going from a $22$ to $21$ layered solid (Fig. 2), and an elastic energy 
cost $\propto \log(l)$ for creating the step,
reveals that a surface step is stable only if 
$l \geq  l^{\ast} \sim 1/L_s$. For small $L_s$, the critical size 
$l^{\ast}$ may therefore  exceed $L_x$, the total length 
of the interface. Indeed, we observe all 
steps, save a complete removed layer, get annealed 
by particles from the adjacent fluid over a time scale of order $\tau$.
% ({\bf What is the $L_s$ dependence ? For macroscopic solids the critical nucleation
%size should be finite, i.e., not scale with $L_x$}).
 The solid therefore relieves stress 
only by the {\it loss or gain of an entire lattice layer}, since all other avenues of 
stress relief entail higher energy costs. \\
%Note that capillary-wave 
%fluctuations of the solid-fluid interface are massive 
%because of the confining potential.
%\vskip .1 cm

\noindent
This mechanism of stress relaxation via the transfer of an entire layer of atoms may be exploited for a 
variety of practical applications, provided we can eject this layer of 
atoms deep into the adjoining fluid and enhance its lifetime. 
Highly stressed monoatomic layers tend to disintegrate or curl up\cite{pnas}
as they separate off from the parent crystal. It may be possible to bypass this 
eventuality, if the time scale of separation is made much smaller than 
the lifetime of the layer. Can acoustic spallation\cite{spall} be used to 
cleave atomic layers from a metastable, stressed nanocrystal?  
Imagine, therefore, sending 
in a sharp laser (or ultrasonic) pulse, producing
a momentum impulse ($v_y(t=0) = V_0$) over a thin region in $y$ spanning the
length $L_x$ of the cell, which results in a weak acoustic 
shock\cite{spall} (corresponding to a laser power $\approx 10^2$ mW and a pulse duration $1 {\rm ps}$ for a typical 
atomic system). 
\begin{figure}[t]
\begin{center}
\includegraphics[width=8.0cm]{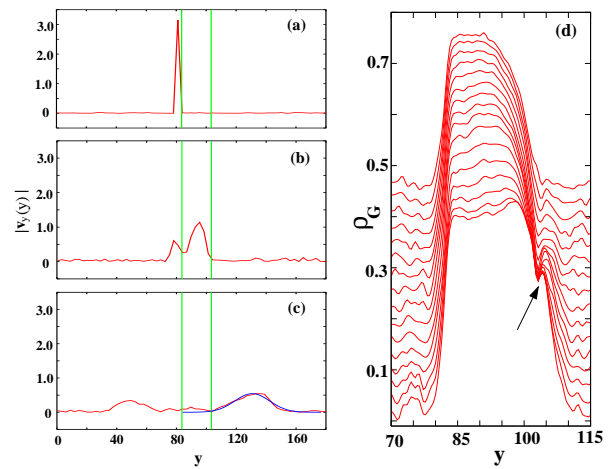}
\end{center}
\caption{
(a)-(c) Plot of the absolute value of the momentum $|v_y(y)|$ for molecular
dynamics times $t = .0007$ (a), $.2828$ (b) and $2.8284$ (c). The 
green lines show the position of the solid~-fluid interfaces. The 
parameters $L_s = 19.73$, $\eta_s = .789$ and $\mu = 4.8$. The fit to a 
Gaussian (blue line) is also shown in (c). The initial momentum pulse 
with strength
$V_0 = 6.$ is given within a narrow strip of size $\sim \sigma$, just to the
left of the solid region and the Gaussian fitted (and the width $\Delta^2$
 extracted) when the maximum of the pulse reaches a fixed distance of $44.1$ 
from the source. A reflected pulse can also be seen. To reduce interference 
from the reflected pulse through periodic boundary 
conditions, we increase the fluid regions on either side, so that for the 
MD calculations we have a cell of size $22.78\times186.98$ comprising 
$3600$ particles. (d) A plot of the time development of the Fourier component 
of the local density correlation $\rho_{\bf G}(y,t)$  
obtained by averaging, at each time slice $t$, the sum 
$\sum_{j=1,N} \exp(-i\,{\bf G\,\cdot}\,({\bf r_j - r_i}))$ over all particle 
positions ${\bf r_i}$ within a strip of width $\sim \sigma$ centered 
about $y$ and spanning the system in $x$. The wavenumber 
${\bf G} = (2 \pi/d) {\bf \hat{n}}$ where
$d = .92$ is the distance between crystal lines in the 
direction ${\bf \hat{n}}$ normal to the fluid-solid interface. The 
solid (central region with $\rho_{\bf G}(y,t) \ne 0$) ejects 
a layer (shown by an arrow) which subsequently dissolves in the fluid. The 
curves from bottom to top correspond to time slices 
at intervals of $\Delta t = .07$ starting from 
$t = 1.06$ (bottom). We have shifted each curve upward by $.03 t/\Delta t$ for 
clarity. Curves such as in (a)-(d) are obtained by averaging 
over $100-300$ separate runs using different realizations of the initial
momentum distribution.}
\label{pulse}
\end{figure}
\vskip .1 cm

\noindent
The initial momentum pulse travels through the solid and 
emerges at the far end (Fig. 3 (a)-(c)) as a broadened Gaussian 
whose width, $\Delta$, is a measure of absorption of the 
acoustic energy of the pulse due to combined dissipation in the liquid, 
the solid and at the interfaces\cite{landau,kapitza}. 
For large enough pulse strengths $V_0$, this is accompanied by a
{\it coherent ejection of the (single) outer layer of atoms into the fluid}; 
such coherent {\it nanospallation} involves surface
stresses of the order $k_B T/\sigma^2 \approx  10^{-5}$ N/cm$^2$ (Fig. 3(d)).
In contrast, spallation in bulk solids
like steel needs acoustic pressures in excess of 
$10^5$ N/cm$^2$\cite{spall} usually available only during impulsive loading
conditions; the ejected layer is a ``chunk'' of the surface. This difference comes about because unlike a bulk system, a strained 
nanocrystal on the verge of a transition from a metastable $n+1$ to a $n$ 
layered state readily absorbs kinetic energy from the pulse. The fact that  
surface indentations are unstable (see Fig. 4(a)) unless of a size comparable to the length of 
the crystal, $L_x$, ensures that a full atomic layer is evicted almost 
always, leading to coherent absorption of the pulse energy.  
The coherence of this absorption 
mechanism is markedly evident in a plot of $\Delta^2$ against $V_0$
which shows a sharp peak (Fig. 4 (b) ). Among the two systems studied by us, {\it viz.}, a 
metastable ($\mu = 4.8$) and a stable ($\mu = 9.6$)  $22$ layered solid,
the former shows a sharper resonance. The eviction of the atomic layer is 
therefore assisted by the strain induced interlayer transition and 
metastability of the $22$ layered solid discussed above. Spallation
is also facillitated if the atomic interactions are anisotropic so that
attraction within layers is stronger than between layers (eg. graphite
and layered oxides\cite{pnas}), for our model, purely repulsive, hard disk
solid, an effective, intralayer attractive potential of mean force is 
induced by the external potential\cite{debc}.
\\
 
\noindent
The spallated solid layer emerges from the solid surface into the fluid, and 
travels 
a distance close to the mean free path; whereupon it disintegrates due 
to viscous dissipation (Fig. 3 (d)). The lifetime of the layer is around 
2-3 time units ($\tau$) which translates to a few ps for typical atomic
systems. The lifetime increases with decreasing viscosity of the surrounding 
fluid. Using the Enskog approximation\cite{enskog} to the hard disk 
viscosity, we estimate that by lowering the fluid density one may increase 
the lifetime by almost three times. The lifetime enhancement is 
even greater if the fluid in contact is a low density gas (when the 
interparticle potential has an attractive part\cite{LJ-visc}). \\
\begin{figure}[t]
\begin{center}
\includegraphics[width=7.5 cm]{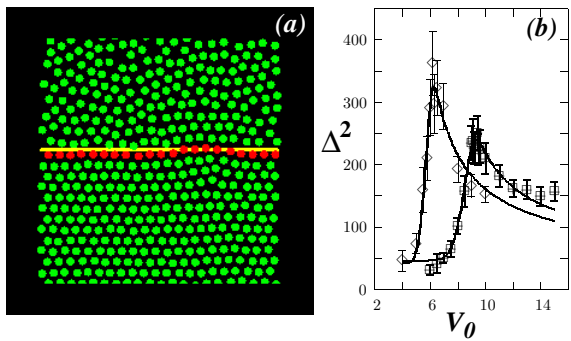}
\end{center}
\caption{(a)(color on-line) 
Configuration snapshot from a portion of our MD cell showing 
hard disk atoms (green circles) at the solid (bottom)- liquid 
interface (yellow line) as a weak momentum pulse ($V_0 = 2.$) emerges 
into the liquid. The pulse initially ejects a few atoms 
of the interfacial crystalline layer (red circles) of a metastable 
$22$ layered solid at $\mu = 4.8$.  
The resulting large non-~uniform elastic strain 
evidenced by the bending of lattice layers, however, causes these atoms 
to be subsequently pulled back into the solid. Only a stronger pulse
capable of ejecting a complete lattice layer succeeds in reducing the 
number of solid layers by one leading to overall lower elastic energy.     
(b) Plot of the squared width $\Delta^2$ of the momentum 
pulse after it emerges from the solid as a function 
of $V_0$ for $\mu = 4.8\,(\,\Diamond)$ and $9.6\,(\,\Box)$.  The 
solid line is a guide to the eye. The absorption of momentum 
is largest when the available kinetic energy of the pulse exactly 
matches the potential energy required to eject a layer. 
The peak in $\Delta^2(V_0)$ so produced is more prominent for 
the metastable $22$ layered solid $\mu = 4.8 $ than for the 
stable ($\mu = 9.6$) system showing a more coherent momentum 
transfer in the former case. 
}
\label{reson}
\end{figure}

\noindent
{\bf Acknowledgements:} The authors thank D. Chaudhuri, G. I. Menon, 
and A. K. Raychaudhuri
for discussions; A. C. thanks C.S.I.R., India, for a fellowship. 
Financial support from DST grant SP/S2/M-20/2001 is gratefully acknowledged.

%%%%%%%%%%%%%%%%%%%%%%%%%%%%%%%%%%%%%%%%%%%%%%%%%%%%%%%%%%%%%%%%%%%%%
%                     REFERENCE PAGE
%%%%%%%%%%%%%%%%%%%%%%%%%%%%%%%%%%%%%%%%%%%%%%%%%%%%%%%%%%%%%%%%%%%%%%


\begin{thebibliography}{}
\bibitem{haasen} Robert W Cahn and Peter Haasen {\em Physical Metallurgy},
(North Holland, Amsterdam, 1996).
\bibitem{marder} J. Fineberg and M. Marder, Phys. Reports, {\bf 313}, 1 
(1999).
\bibitem{uzi} U. Landman et al., Science {\bf 248}, 454 (1990);
A.P. Sutton and J.B. Pethica, J. Phys. Condens. Matt. {\bf 2}, 5317 (1990).
\bibitem{debc}
D. Chaudhuri and S. Sengupta, \prl {\bf 93}, 115702 (2004).
\bibitem{daan}
D. Frenkel and B. Smit {\em Understanding Molecular Simulations},
(North Holland, Amsterdam, 2000)
\bibitem{landau-el}
L. D. Landau and E. M. Lifshitz {\em Theory of Elasticity, 2$^{nd}$ Edition}
(Pergamon Press, Oxford, 1987).
\bibitem{landau}
L. D. Landau and E. M. Lifshitz {\em Fluid Mechanics, 2$^{nd}$ Edition}
(Pergamon Press, Oxford, 1987).
C. Zener, Phys. Rev. {\bf 53}, 90, (1938). 
\bibitem{laser-trp}
H. J. Metcalf and P. van der Straten {\em Laser Cooling and Trapping}
(Springer, Heidelberg 1999)
\bibitem{spall}
Ya. B. Zel'dovich and Yu. P. Raizer {\em Physics of Shock Waves and High-Temperature Hydrodynamic Phenomena} (Dover Publications, New York, 2002).
\bibitem{kapitza}
D. G. Cahill {\em et. al.}, J. Appl. Phys. {\bf 93}, 793, (2003) and 
references therein. 
\bibitem{nanostuff}
C. P. Poole and F. J. Owens {\em Introduction to nanotechnology}, 
(Wiley, New Jersey, 2003). 
\bibitem{nems}
V. Balzani, M. Venturi and A. Credi, 
{\em Molecular devices and machines 
: a journey into the nano world
}
(Wiley-VCH, Weinheim, 2003).
\bibitem{pnas} K. S. Novoselov {\em et. al.}, Proc. Nat. Acad. Sc. {\bf 102},
10451 (2005).
\bibitem{sakhi} S. Psakhie, Bull. Amer. Phys. Soc. {\bf 46}, C3.005, (2001).
\bibitem{colbook}
I. W. Hamley {\em Introduction to Soft Matter: polymer, colloids, amphiphiles
and liquid crystals} (Wiley, Cluchester, 2000).
\bibitem{col-trp}
D. G. Grier, Nature, {\bf 424}, 810 (2003).
\bibitem{laser-col}
J. Baumgartl, M. Brunner, and C. Bechinger,
\prl {\bf 93}, 168301 (2004).
\bibitem{epitaxy}
J. P. Hoogenboom {\em et. al.}, Appl. Phys. Lett. {\bf 80}, 4828 (2002).
\bibitem{jaster}
A. Jaster, Physica A.{\bf 277}, 106 (2000);
\bibitem{note}Typical equilibration times are large and so we discard many ($\sim 10^7$) Monte Carlo steps (MCS) before computing equilibrium averages. 
\bibitem{snb}
S. Sengupta, P. Nielaba, K. Binder, Phys. Rev. {\bf E 61}, 6294 (2000).
\bibitem{frevol}
M. Heni and H. L\"owen, Phys. Rev.  {\bf E 60}, 7057 (1999).
\bibitem{santos}
A. Santos, M. Lo´pez de Haro and S. Bravo Yuste,
J. Chem. Phys. {\bf 103}, 4622, (1995).
\bibitem{note-energy}A time step of $\Delta t = 10^{-4}$
conserves the total energy to within $1$ in $10^{3}$(at worst)$-10^{6}$.
\bibitem{enskog} S. Chapman and T. G. Cowling, {\em The Mathematical Theory
of Non-uniform Gases, 3$^{rd}$ Ed.}, (Cambridge University Press, London
1970).
\bibitem{LJ-visc} W. G. Hoover {\em et. al.},
Phys. Rev. A, {\bf 22}, 1690 (1980).

\end{thebibliography}
\end{document}